\documentclass[a4paper,12pt]{article}

\usepackage{graphics, epsfig}

\begin{document}

\author{C. Barrab\`es\thanks{E-mail : barrabes@celfi.phys.univ-tours.fr}\\
\small Laboratoire de Math\'ematiques et Physique Th\'eorique,\\
\small CNRS/UMR 6083, Universit\'e F. Rabelais, 37200 TOURS,
France\\\small and \\P. A. Hogan\thanks{E-mail : peter.hogan@ucd.ie}\\
\small School of Physics,\\ \small University College Dublin,
Belfield, Dublin 4, Ireland}

\title{Inhomogeneous High Frequency Expansion--Free Gravitational Waves}
\date{(To appear in Physical Review D)}
\maketitle

\begin{abstract}
We describe a natural inhomogeneous generalization of high
frequency plane gravitational waves. The waves are high frequency
waves of the Kundt type whose null propagation direction in
space--time has vanishing expansion, twist and shear but is not
covariantly constant. The introduction of a cosmological constant is
discussed in some detail and a comparison is made with high frequency
gravity waves having wave fronts homeomorphic to 2--spheres.
\end{abstract}
\thispagestyle{empty}
\newpage

\section{Introduction}\indent
Two important special families of gravitational waves, from the
point of view of modelling astrophysical processes, are bursts of
gravitational radiation (which may be accompanied by matter travelling with the speed of light) and
high frequency gravitational waves. For an overview of the former see \cite{BH}. This
paper is concerned with establishing the line--elements of the space--time
models of the gravitational fields of high frequency gravitational waves. For simplicity
we consider monochromatic waves here. These
are the basic building blocks for the study of the interaction of such waves in
General Relativity and for constructing model systems responsible for such waves.
They could be used to further develop the interesting ideas on the self--interaction
of gravitational waves in \cite{SMDMC} and \cite{MV} for example.

Einstein's theory of General Relativity is similar to
Maxwell's theory of Electromagnetism in permitting plane wave
solutions which are both homogeneous and inhomogeneous. Einstein's
theory does not however permit exactly spherical waves (on account
of the Birkhoff theorem) but it does of course permit waves from isolated
sources having wave fronts which are homeomorphic to 2--spheres. A
physically important subfield of the theory of waves in General
Relativity is concerned with modelling the gravitational fields
due to high frequency or short wavelength gravitational waves. The
generation of such waves by a high frequency compact source in the field
of a black hole, for example, and the influence of the background curvature
on the wave propagation is studied in \cite{PPK} (for an alternative
mechanism generating such waves see \cite{PL}--\cite{SO3}; for a complementary
approach see Ellis \cite{E}).
Homogeneous, plane, high frequency gravitational waves are well known
\cite{Burnett} and so this paper is primarily concerned with
providing examples of high frequency, \emph{inhomogeneous}, non--expanding
gravitational waves and comparing these with the corresponding
waves having roughly spherical wave fronts. The paper is organised
as follows: In section 2 high frequency, homogeneous plane waves
are described as a means of introducing our point of view. The
generalization to \emph{inhomogeneous} waves is given in
section 3. If the inhomogeneous waves of section 3 are $pp$--waves
(plane fronted waves with parallel rays) then the description of their
high frequency version in the presence of a cosmological constant is
given in section 4. The resulting high frequency waves are not pp--waves
since their propagation direction in space--time is no longer covariantly
constant in the presence of a cosmological constant. This is followed in section 5 by a comparison with
high frequency `spherical' waves. To make the paper as self contained as
possible we provide calculational details in appendices.

\setcounter{equation}{0}
\section{Homogeneous Waves}\indent
The line--element of the space--time model of the (approximately)
vacuum gravitational field due to a train of short wavelength or
high frequency homogeneous, plane gravitational waves can be
written in the form \cite{Burnett}
\begin{equation}\label{1.1}
ds^2=-2\,P_{\lambda}(u)^{-2}\,| d\zeta +\lambda\,\bar
W(u)\,\sin\frac{u}{\lambda}\,d\bar\zeta |^2+2\,du\,dr\
.\end{equation}Here $\zeta$ is a complex coordinate with complex
conjugate here and throughout indicated by a bar. The coordinates
$r, u$ are real. $P_{\lambda}$ is a real--valued function of the
coordinate $u$ with a dependence on the real parameter
$\lambda\geq 0$ on account of Einstein's vacuum field equations.
$W$ is an arbitrary complex--valued function of the real
coordinate $u$. The hypersurfaces $u={\rm constant}$ are null and
are generated by the null geodesic integral curves of the vector
field $k^a\partial /\partial x^a=\partial /\partial r$ with $r$
and affine parameter along them. This vector is in fact
covariantly constant and the null hypersurfaces are the histories
of the plane wave fronts of the gravitational waves. The parameter
$\lambda$ will play the role of the wavelength of the
gravitational waves and we shall assume that $\lambda$ is small.
For small $\lambda$ the vacuum field equations are approximately
satisfied by the metric tensor given via the line-element
(\ref{1.1}) in the sense that the Ricci tensor satisfies
\begin{equation}\label{1.2}
R_{ab}=O(\lambda )\ ,\end{equation}provided $P_{\lambda}(u)$
satisfies
\begin{equation}\label{1.3}
-\dot H_{\lambda}+H^2_{\lambda}+|W|^2\sin ^2\frac{u}{\lambda}=0\
,\end{equation}where the dot indicates differentiation with
respect to $u$ and
\begin{equation}\label{1.4}
H_{\lambda}=P_{\lambda}^{-1}\dot P_{\lambda}\ .\end{equation}The
components of the Riemann curvature tensor in Newman--Penrose
notation are denoted by $\Psi _A$ with $A=0, 1, 2, 3, 4$ and when
calculated with the metric tensor given via (\ref{1.1}) we find
that $\Psi _A=O(\lambda )$ for $A=0, 1, 2, 3$ and
\begin{equation}\label{1.5}
\Psi _4=\lambda ^{-1}W\,\sin\frac{u}{\lambda}+O(\lambda ^0)\
.\end{equation}Thus for small $\lambda$ the Riemann tensor is type
N in the Petrov classification having $\partial /\partial r$ as
degenerate principal null direction. Equation (\ref{1.5})
indicates a profile for these waves which has large amplitude and
short wavelength. Using the Riemann--Lebesgue theorem
\cite{Burnett} (described in Appendix A) one can deduce from
(\ref{1.3}) that for small $\lambda$
\begin{equation}\label{1.6}
P_{\lambda}=P_0+\frac{1}{8}\lambda
^2P_0|W|^2\cos\frac{2\,u}{\lambda}+O(\lambda ^3)\
,\end{equation}with $P_0=\lim _{\lambda\rightarrow 0}P_{\lambda}$
satisfying the differential equation
\begin{equation}\label{1.7}
-\dot H_0+H^2_0+\frac{1}{2}\,|W|^2=0\ ,\end{equation}with
$H_0=P_0^{-1}\dot P_0$. We observe that there is no $O(\lambda )$--term in
(\ref{1.6}) but the $O(\lambda ^2)$--term is necessary in order to
satisfy (\ref{1.2}) and thus (\ref{1.3}). Now the line--element (\ref{1.1})
naturally splits, for small $\lambda$, into the form
\begin{equation}\label{1.8}
ds^2=d\hat s^2+O(\lambda )\ ,\end{equation}with
\begin{equation}\label{1.9}
d\hat s^2=-2\,P_0^{-2}\,| d\zeta |^2+2\,du\,dr\ .\end{equation}This
`background' line--element has Ricci tensor
\begin{equation}\label{1.10}
\hat R_{ab}=-|W|^2k_a\,k_b\ ,\end{equation}on account of
(\ref{1.7}). We note that $k^a$ given following (\ref{1.1}) above
is equivalently given by the 1--form $k_a\,dx^a=du$. The lambda
dependence in the equations (\ref{1.2}) and (\ref{1.5}) and the
algebraic form of the background Ricci tensor (\ref{1.10}) are
what one expects in general for high frequency gravitational waves
following the pioneering work of Isaacson \cite{I1}\cite{I2} (see
also \cite{CB}, \cite{MCT}). The coordinates $\zeta , \bar\zeta$
are intrinsic to the histories of the plane wave fronts and the
absence of the dependence of any functions on these coordinates in
the example given here reflects the fact that the waves here are
\emph{homogeneous}.

\setcounter{equation}{0}
\section{Inhomogeneous Generalization}\indent
The generalization to \emph{inhomogeneous}, high frequency non--expanding
waves which we wish to consider here is expressed by the
line--element:
\begin{equation}\label{2.1}
ds^2=-2\,P_{\lambda}^{-2}\,|d\zeta +\lambda\,P_{\lambda}^2\bar
W(\bar\zeta , u)\,\sin\frac{u}{\lambda}\,d\bar\zeta
|^2+2\,du\,dr+(S_{\lambda}-2\,r\,H_{\lambda})\,du^2\
.\end{equation}Here $P_{\lambda}(\zeta , \bar\zeta , u)$ is a
real--valued function satisfiying
\begin{equation}\label{2.2}
\Delta _{\lambda}\log P_{\lambda}=0\ ,\qquad \Delta
_{\lambda}=2\,P_{\lambda}^2\,\frac{\partial
^2}{\partial\zeta\partial\bar\zeta}\ ,\end{equation}and
\begin{equation}\label{2.3}
H_{\lambda}=P_{\lambda}^{-1}\dot  P_{\lambda}\ .\end{equation}Also
$W(\zeta , u)$ is an arbitrary analytic function of its arguments
while $S_{\lambda}(\zeta , \bar\zeta , u)$ is a real--valued
function satisfying
\begin{equation}\label{2.4}
-\dot H_{\lambda}+2\,H_{\lambda}^2-\frac{1}{4}\Delta
_{\lambda}S_{\lambda}+P_{\lambda}^4|W|^2\sin
^2\frac{u}{\lambda}=0\ .\end{equation}Equation (\ref{2.4}) is the
analogue here of (\ref{1.3}) and ensures that the metric given via
the line--element (\ref{2.1}) satisfies (\ref{1.2}). Using the
Riemann--Lebesgue theorem \cite{Burnett} (see Appendix A) one
deduces from (\ref{2.4}) that for small $\lambda$
\begin{equation}\label{2.5}
P_{\lambda}=P_0+\frac{1}{8}\lambda
^2P_0^5|W|^2\cos\frac{2\,u}{\lambda}+O(\lambda ^3)\
,\end{equation}with $P_0$ satisfying the differential equation
\begin{equation}\label{2.6}
-\dot H_0+2\,H^2_0-\frac{1}{4}\Delta
_0S_0+\frac{1}{2}P_0^4|W|^2=0\ ,\end{equation}and the subscript
zero denoting that the limit $\lambda\rightarrow 0$ has been
taken. The Newman--Penrose components of the Riemann curvature
tensor calculated with the metric tensor given via (\ref{2.1}) now
satisfy $\Psi _A=O(\lambda )$ for $A=0, 1, 2$ and
\begin{eqnarray}\label{2.7}
\Psi _3&=&-P_0\,\frac{\partial
H_0}{\partial\zeta}+O(\lambda )\ ,\\
\Psi _4&=&\lambda ^{-1}P_0^2W\,\sin\frac{u}{\lambda}+O(\lambda
^0)\ .\end{eqnarray}Thus for small $\lambda$ the space--time with
line--element (\ref{2.1}) is a vacuum space--time (on account of
(\ref{1.2})) and is type N in the Petrov classification (since for
small $\lambda$ the Riemann curvature tensor is dominated by the
Newman--Penrose component $\Psi _4$) and thus is a model of the
gravitational field of low wavelength, high amplitude
gravitational waves. These waves are inhomogeneous plane waves in
the sense of Kundt \cite{K} because their null propagation
direction in space--time $k^a\partial /\partial x^a=\partial
/\partial r$ is twist--free, shear--free and expansion--free. They
are in fact a subclass of such space--times because the covariant
null vector field $k_a\,dx^a=du$ in the space--time with
line--element (\ref{2.1}) has covariant derivative, with respect
to the Riemannian connection calculated with the metric tensor
given via (\ref{2.1})(and indicated by a stroke), of the form
\begin{equation}\label{2.8}
k_{a|b}=H_{\lambda}\,k_a\,k_b\ .\end{equation}Vacuum space--times
admitting a null vector field satisfying an equation of this form
were first discussed in detail in \cite{DC} and \cite{GK}. We see from
(\ref{2.2}) that $P_{\lambda}=|h_{\lambda}(\zeta , u)|$ for some analytic
function $h_{\lambda}(\zeta , u)$. The special case $P_0=1$ can be realized
with $h_{\lambda}(\zeta , u)=1+O(\lambda ^2)$ on account of (\ref{2.5}). This
leads to $H_{\lambda}=O(\lambda )$ and $\dot H_{\lambda}=O(\lambda ^0)$ and
so (\ref{2.8}) reads $k_{a|b}=O(\lambda )$. Hence for small $\lambda$ the inhomogeneous
high frequency waves in this case are the so--called $pp$--waves \cite{MaC}
(plane--fronted waves with parallel rays). We
will consider this special case in section 4 in order to describe the
introduction of a cosmological constant. The introduction of a cosmological
constant in the general case in which (\ref{2.8}) holds is an interesting
nontrivial topic for further study.

The line--element (\ref{2.1}) can be written, for small $\lambda$,
as a background plus a small perturbation as in (\ref{1.8}) but
now with
\begin{equation}\label{2.9}
d\hat s^2=-2\,P_0^{-2}\,|d\zeta |^2+2\,du\,dr+(S_0-2\,r\,H_0)\,du^2\
.\end{equation}The Ricci tensor calculated with the metric given
by this line--element is given by
\begin{equation}\label{2.10}
\hat R_{ab}=-P_0^4|W|^2\,k_a\,k_b\ ,\end{equation}on account of
(\ref{2.6}). Thus all of the classical ingredients of high
frequency gravity waves described at the end of section 2 are
satisfied by these inhomogeneous plane waves for small $\lambda$.

\setcounter{equation}{0}
\section{Expansion--free Waves with Cosmological Constant}\indent
The classic study of expansion--free gravitational waves with a cosmological
constant is the paper by Ozsv\'ath, Robinson and R\'ozga \cite{ORR} (for studies
of these space--times and their geometric subclasses see \cite{GP}--\cite{BP}). We
will rely heavily on this work in our discussion of high frequency waves of
this type. The line--element of interest to us is a generalization of (\ref{2.1})
given by
\begin{equation}\label{4.1}
ds^2=-2P_{\lambda}^{-2}\,\left |d\zeta +\lambda\,P_{\lambda}^2\bar W(\bar\zeta , u)\,
\sin\frac{u}{\lambda}\,d\bar\zeta\right |^2+2\,B_{\lambda}\,du\,(dr+\frac{1}{2}c_{\lambda}\,
du)\ ,\end{equation}with $P_{\lambda}\,, B_{\lambda}$ real--valued functions
of $\zeta , \bar\zeta$ and $u$, $c_{\lambda}$ a real--valued function
of $\zeta , \bar\zeta ,\ r$ and $u$ and $W(\zeta , u)$ a complex--valued analytic function of $\zeta$ and
 $u$. For $\lambda >0$ small we require the metric tensor given by this line--element to
 satisfy approximately Einstein's vacuum field equations with a cosmological constant $\Lambda$:
 \begin{equation}\label{4.2}
 R_{ab}=\Lambda\,g_{ab}+O(\lambda )\ .\end{equation}The detailed results of the calculations
 of the Ricci tensor here are listed for convenience in appendix B. Together with (\ref{4.2})
 they give us the following:
 \begin{equation}\label{4.3}
 \frac{\partial}{\partial\zeta}\left (P_{\lambda}^2\,
 \frac{\partial B_{\lambda}^{1/2}}{\partial\zeta}\right )=0\ ,\end{equation}
and
\begin{equation}\label{4.4}
\Delta _{\lambda}\log P_{\lambda}-\frac{1}{2}B_{\lambda}^{-1}\,\Delta _{\lambda}
B_{\lambda}+\frac{1}{2}B_{\lambda}^{-2}P_{\lambda}^2\left |\frac{\partial B_{\lambda}}
{\partial\zeta}\right |^2=\Lambda\ .\end{equation}The function $c_{\lambda}$ takes the form
\begin{equation}\label{4.5}
c_{\lambda}=h_{\lambda}(u)\,r^2+f_{\lambda}(\zeta , \bar\zeta , u)\,r+g_{\lambda}(\zeta , \bar
\zeta ,u)\ ,\end{equation}with
\begin{eqnarray}\label{4.6}
h_{\lambda}(u)&=&-B_{\lambda}\,\Lambda -\frac{1}{2}\Delta _{\lambda}B_{\lambda}\ ,\\
\frac{\partial f_{\lambda}}{\partial\zeta}&=&-2\,\frac{\partial H_{\lambda}}{\partial\zeta}
+\frac{\partial ^2}{\partial u
\partial\zeta}(\log B_{\lambda})+2\,H_{\lambda}\,\frac{\partial}{\partial\zeta}(\log B_{\lambda})\ ,\end{eqnarray}
\begin{equation}\label{4.7}
\Delta _{\lambda}f_{\lambda}+P_{\lambda}^2\left (\frac{\partial}{\partial\bar\zeta}(\log B_{\lambda})\,
\frac{\partial f_{\lambda}}{\partial\zeta}+\frac{\partial}{\partial\zeta}(\log B_{\lambda})\,
\frac{\partial f_{\lambda}}{\partial\bar\zeta}\right )+4\,B_{\lambda}^{-1}H_{\lambda}\,h_{\lambda}=0\ ,
\end{equation}and
\begin{eqnarray}\label{4.8}
-\frac{1}{4}\Delta _{\lambda}g_{\lambda}&-&\frac{1}{4}P_{\lambda}^2B_{\lambda}^{-1}
\left (\frac{\partial B_{\lambda}}{\partial\bar\zeta}\,\frac{\partial g_{\lambda}}{\partial\zeta}+
\frac{\partial B_{\lambda}}{\partial\zeta}\,\frac{\partial g_{\lambda}}{\partial\bar\zeta}\right )-
\frac{1}{2}B_{\lambda}^{-1}H_{\lambda}\,f_{\lambda}\nonumber \\
&-&B_{\lambda}^{-1}\dot H_{\lambda}+
B_{\lambda}^{-1}H_{\lambda}^2+B_{\lambda}^{-2}\dot B_{\lambda}\,H_{\lambda}+B_{\lambda}^{-1}
P_{\lambda}^4|W|^2\sin ^2\frac{u}{\lambda}\nonumber \\
&=&0\ .\end{eqnarray}

Differentiating (\ref{4.4}) with respect to $u$ results in
\begin{equation}\label{4.9}
\Delta _{\lambda}H_{\lambda}+2\,\Lambda\,H_{\lambda}=P_{\lambda}^2\frac{\partial}{\partial u}\left (
B_{\lambda}^{-1}\frac{\partial ^2B_{\lambda}}{\partial\zeta\partial\bar\zeta}-\frac{1}{2}\,
B_{\lambda}^{-2}\frac{\partial B_{\lambda}}{\partial\zeta}\,\frac{\partial B_{\lambda}}{\partial\bar\zeta}
\right )\ ,\end{equation}while the integrability of (4.7) gives us
\begin{equation}\label{4.10}
\frac{\partial B_{\lambda}}{\partial\zeta}\,\frac{\partial H_{\lambda}}{\partial\bar\zeta}=
\frac{\partial B_{\lambda}}{\partial\bar\zeta}\,\frac{\partial H_{\lambda}}{\partial\zeta}\ .\end{equation}
Now substituting for $h_{\lambda}(u)$ from (\ref{4.6}) and for $\partial f_{\lambda}/\partial\zeta$ from
(4.7) into (\ref{4.7}) shows that (\ref{4.7}) is automatically satisfied on account of (\ref{4.9})
and (\ref{4.10}). Before returning to (\ref{4.10}) we begin by solving (\ref{4.4}). We are initially
guided by the special case $\Lambda =0$ and the expression for $P_{\lambda}$ in that case
given in the discussion following (\ref{2.8}), and also by the results in \cite{ORR}. We take
\begin{equation}\label{4.11}
P_{\lambda}=\left (1+\frac{\Lambda}{6}\,|w_{\lambda}|^2\right )\,|w'_{\lambda}|^{-1}\ ,
\end{equation}where $w_{\lambda}(\zeta , u)$ is an analytic function and $w_{\lambda}'=\partial w_{\lambda}/\partial\zeta$. In addition
we shall assume that $B_{\lambda}=B_{\lambda}(w_{\lambda}, \bar w_{\lambda}, u)$. Now (\ref{4.3}) reads
\begin{equation}\label{4.12}
\frac{\partial}{\partial w_{\lambda}}\left (\left (1+\frac{\Lambda}{6}\,|w_{\lambda}|^2\right )
^2\frac{\partial B_{\lambda}^{1/2}}{\partial w_{\lambda}}\right )=0\ ,\end{equation}
while (\ref{4.4}) takes the form
\begin{equation}\label{4.13}
\left (1+\frac{\Lambda}{6}\,|w_{\lambda}|^2\right )^2\,\frac{\partial ^2B_{\lambda}^{1/2}}
{\partial w_{\lambda}\partial\bar w_{\lambda}}+\frac{1}{3}\,\Lambda\,B_{\lambda}^{1/2}=0\ .\end{equation}
These equations are satisfied by the Ozsv\'ath-Robinson-R\'ozga \cite{ORR} function
\begin{equation}\label{4.14}
B_{\lambda}^{1/2}=\left (1+\frac{\Lambda}{6}\,|w_{\lambda}|^2\right )^{-1}
\left\{\alpha (u)\left (1-\frac{\Lambda}{6}\,|w_{\lambda}|^2\right )+\beta (u)\,\bar w_{\lambda}+
\bar\beta (u)\,w_{\lambda}\right\}\ ,\end{equation}with $\alpha$ an arbitrary real--valued function of
$u$ and $\beta$ an arbitrary complex--valued function of $u$. Substitution into (4.6) yields
\begin{equation}\label{4.15}
h_{\lambda}=-\frac{1}{3}\alpha ^2\Lambda -2\,\beta\,\bar\beta =-\kappa (u)\ ({\rm say})\ .\end{equation}
Referring to the discussion of the function $P_{\lambda}(\zeta , \bar\zeta , u)$ following (\ref{2.8}) we
can obtain the corresponding waves here with $\Lambda\neq 0$ by assuming that for small $\lambda >0$ we
can write
\begin{equation}\label{4.16}
w_{\lambda}(\zeta , u)=\zeta +\lambda ^2g_2\left (\frac{u}{\lambda}\right )\,{\cal G}_2(\zeta , u)+\dots
\ .\end{equation}It follows from this that $H_{\lambda}=O(\lambda )$ and $\dot H_{\lambda}=O(\lambda ^0)$.
This circumvents the integrability condition (\ref{4.10}) because the field equation (4.7) is
now replaced by
\begin{equation}\label{4.17}
\frac{\partial f_{\lambda}}{\partial\zeta}=\frac{\partial ^2}{\partial\zeta\partial u}(\log B_{\lambda})\ .
\end{equation}We note that $O(\lambda )$--terms are neglected in all field equations as a consequence of
the assumption (\ref{4.2}). We take as solution of (\ref{4.17}) the function
\begin{equation}\label{4.18}
f_{\lambda}=\frac{\partial}{\partial u}(\log B_{\lambda})\ .\end{equation} Now by (\ref{4.11}) we have
\begin{equation}\label{4.19}
P_{\lambda}=p+O(\lambda ^2)\ ,\qquad p=1+\frac{\Lambda}{6}\zeta\bar\zeta\ ,
\end{equation}and by (\ref{4.14}),
\begin{equation}\label{4.20}
B_{\lambda}^{1/2}=p^{-1}q+O(\lambda ^2)\ ,\qquad q=\alpha\,\left (1-\frac{\Lambda}{6}\zeta\bar\zeta
\right )+\beta\,\bar\zeta +\bar\beta\,\zeta\ .\end{equation}The final field equation (always neglecting
$O(\lambda )$--terms) is (\ref{4.8}). Writing $S_{\lambda}=B_{\lambda}^{1/2}g_{\lambda}=p^{-1}q\,g_{\lambda}$
(approximately), this last field equation becomes:
\begin{equation}\label{4.21}
-\frac{1}{4}p^{-1}q\,\left (\Delta S_{\lambda}+\frac{2}{3}\,\Lambda\,S_{\lambda}\right )-\dot H_{\lambda}
+p^4|W|^2\sin ^2\frac{u}{\lambda}=0\ ,\end{equation}with
\begin{equation}\label{4.22}
\Delta S_{\lambda}=2\,p^2\,\frac{\partial ^2\,S_{\lambda}}{\partial\zeta\partial
\bar\zeta}\ .\end{equation}Using the argument outlined in Appendix A we find that (\ref{4.19})
can be made more accurate to read
\begin{equation}\label{4.23}
P_{\lambda}=p+\frac{\lambda ^2}{8}\,p^5|W|^2\cos\frac{2u}{\lambda}+O(\lambda ^2)\ ,\end{equation}
and $S_0=\lim_{\lambda\rightarrow 0}S_{\lambda}$ satisfies the differential equation
\begin{equation}\label{4.24}
\frac{\partial ^2S_0}{\partial\zeta\partial\bar\zeta}+\frac{1}{3}p^{-2}\Lambda\,S_0=p^3q^{-1}|W|^2\ .\end{equation}
The Weyl tensor of the space--time under consideration has Newman--Penrose components $\Psi_A=O(\lambda ^0)$
for $A=0, 1, 2, 3$ and
\begin{equation}\label{4.25}
\Psi _4=\lambda ^{-1}p^4q^{-2}W\,\sin\frac{u}{\lambda}+O(\lambda ^0)\ ,\end{equation}which is therefore
predominantly the radiative type with propagation direction given by the 1--form $k_a\,dx^a=du$.

Now the line--element (\ref{4.1}) subdivides into a background and an $O(\lambda )$--term with the
background given by
\begin{equation}\label{4.26}
d\hat s^2=-2\,p^{-2}\,|d\zeta |^2+2\,p^{-2}q^2du\,\left\{dr+\frac{1}{2}\left (p\,q^{-1}S_0+2\,q^{-1}\dot q\,r
-\kappa (u)\,r^2\right )\,du\right\}\ ,\end{equation}and the Ricci tensor of this space--time takes the form
\begin{equation}\label{4.27}
\hat R_{ab}=\Lambda\,g_{ab}-p^4\,|W|^2k_a\,k_b\ .\end{equation} Removing the high--frequency waves (
by putting $W=0$) results in this line--element coinciding with the Ozsv\'ath--Robinson--R\'ozga line--
element \cite{ORR}. The function $S_0$, in this case satisfying (\ref{4.24}) with $W=0$, is required
for the description of waves which include generalized Kundt waves. If $S_0=0$ in (\ref{4.26}) then
this is the line--element of the de Sitter universe in a coordinate system based on a family of null
hypersurfaces $u={\rm constant}$ whose generators have vanishing expansion and shear (see Appendix C).

\setcounter{equation}{0}
\section{Comparison with `Spherical' Waves}\indent

The gravitational waves emitted from isolated gravitating systems
are asymptotically almost spherical. The wave fronts have
histories in space--time which are expanding, shear--free null
hypersurfaces. The space--time models of the vacuum gravitational
fields of such waves in the high frequency approximation, including a
cosmological constant $\Lambda$, have line--elements which are roughly similar to (\ref{2.1}).
Specifically they are given by \cite{FH}
\begin{equation}\label{3.1}
ds^2=-2\,r^2p_{\lambda}^{-2}\,\left |d\zeta
+\frac{\lambda\,p_{\lambda}^2}{r}\,\bar W(\bar\zeta
,u)\sin\frac{u}{\lambda}\,d\bar\zeta\right
|^2+2\,du\,dr+c_{\lambda}\,du^2\ .\end{equation}Here
$p_{\lambda}(\zeta , \bar\zeta , u)$ is a real--valued function,
$W$ is an arbitrary analytic function and the real--valued
function $c_{\lambda}(\zeta , \bar\zeta , u, r)$ is given by
\begin{equation}\label{3.2}
c_{\lambda}=K_{\lambda}-2\,r\,H_{\lambda}-\frac{1}{3}\Lambda\,r^2-\frac{2\,m_{\lambda}}{r}\
,\end{equation}where
\begin{equation}\label{3.3}
K_{\lambda}=\Delta _{\lambda}\log P_{\lambda}\ ,\qquad
H_{\lambda}=P_{\lambda}^{-1}\dot P_{\lambda}\ ,\end{equation}with
\begin{equation}\label{3.4}
\Delta _{\lambda}=2\,p_{\lambda}^{2}\,\frac{\partial
^2}{\partial\zeta\partial\bar\zeta}\ ,\qquad
m_{\lambda}=m_{\lambda }(u)\ .\end{equation}Once again the Ricci
tensor calculated with the metric given by (\ref{3.1}) has the
form (\ref{4.2}) for small $\lambda$ provided the following field
equation is satisfied:
\begin{equation}\label{3.5}
\dot m_{\lambda}-3\,m_{\lambda}\,H_{\lambda}-\frac{1}{4}\Delta
_{\lambda}K_{\lambda}+p_{\lambda}^4|W|^2\sin
^2\frac{u}{\lambda}=0\ .\end{equation}The propagation direction
(in the space--time with line--element (\ref{3.1})) of the
histories of the wave fronts is $k^a\partial /\partial
x^a=\partial /\partial r$ and the integral curves of this vector
field have real expansion $r^{-1}$ and complex shear given by
\begin{equation}\label{3.6'}
\sigma
=\frac{\lambda\,p_{\lambda}^2}{r^2}\,W\,\sin\frac{u}{\lambda}+O(\lambda
^2)=O(\lambda )\ .\end{equation}Thus in the limit
$\lambda\rightarrow 0$ these are expanding, shear--free null
geodesics.

The background (corresponding to $\lambda =0$) Ricci tensor is now
given by
\begin{equation}\label{3.6}
\hat R_{ab}=-r^{-2}p_0^4\,|W|^2\,k_a\,k_b+\Lambda\,\hat g_{ab}\ ,\end{equation}with $p_0$
(and $m_0$) satisfying the equation (see Appendix A)
\begin{equation}\label{3.7}
\dot m_0-3\,m_0\,H_0-\frac{1}{4}\Delta
_0K_0+\frac{1}{2}p_0^4|W|^2=0\ ,\end{equation} and the background
line--element $d\hat s^2=\hat g_{ab}\,dx^a\,dx^b$ has Robinson--Trautman \cite{RT} form.
For small $\lambda >0$ the Riemann curvature
tensor calculated with the metric given by the line--element
(\ref{3.1}) is now $\Psi _A=O(\lambda ^0)$ for $A=0, 1, 2, 3$ and
\begin{equation}\label{3.8}
\Psi _4=\frac{1}{r}\lambda
^{-1}p_0^2W\,\sin\frac{u}{\lambda}+O(\lambda ^0)\ .\end{equation}
To obtain an equation for $p_{\lambda}$ which corresponds to
(\ref{1.6}) and (\ref{2.5}) we first note that the form of the
line--element (\ref{3.1}) is preserved with an $O(\lambda
^2)$--error (sufficient for our purposes) by a coordinate
transformation of the form (a special case of a Robinson--Trautman
\cite{RT} transformation)
\begin{equation}\label{3.9}
u=u'+\lambda ^2w_2\left (\frac{u'}{\lambda}\right )\,\gamma
(u')+O(\lambda ^3)\ ,\qquad r=\frac{du}{du'}\,r'\ ,
\end{equation}for some real--valued functions
$w_2$ and $\gamma$. Under such a transformation the function
$m_{\lambda}$ is transformed to
\begin{equation}\label{3.10}
\hat m_{\lambda}=(1+3\,\lambda\,w'_{2}\,\gamma +O(\lambda
^2))\,m_{\lambda}\ ,\end{equation}with the prime on $w_2$ denoting
differentiation with respect to its argument. For small $\lambda
>0$ we can write (see remark in Appendix A and (\ref{A2}))
\begin{eqnarray}\label{3.11}
m_{\lambda}(u)&=&m_0(u)+\lambda\,q_1\left (\frac{u}{\lambda}\right
)\,m_1(u)+O(\lambda ^2)\ ,\nonumber\\
&=&m_0(u')+\lambda\,q_1\left (\frac{u'}{\lambda}\right
)\,m_1(u')+O(\lambda ^2)\ .\end{eqnarray}Substituting (\ref{3.11})
into (\ref{3.10}) we see that we can choose the functions $w_2\,,\
\gamma$ in the coordinate transformation  (\ref{3.9}) so as to
eliminate the $O(\lambda )$--term in $\hat m_{\lambda}$. Thus
without loss of generality we may take
\begin{equation}\label{3.12}
m_{\lambda}(u)=m_0(u)+O(\lambda ^2)\ ,\end{equation}in
(\ref{3.2}), (\ref{3.4}) and (\ref{3.5}). Now solving (\ref{3.5})
using (\ref{3.7}) yields
\begin{equation}\label{3.13}
p_{\lambda}=p_0-\frac{1}{12}\,\lambda\,m_0^{-1}p_0^5\,|W|^2\sin\frac{2\,u}{\lambda}
+O(\lambda ^2)\ .\end{equation}

\setcounter{equation}{0}
\section{Conclusion}\indent
The pattern of the line--elements (\ref{1.1}), (\ref{2.1}),
(\ref{4.1}) and (\ref{3.1}) is now clear and they could reasonably claim to be
the basic non--expanding and expanding high frequency gravitational wave solutions of
Einstein's field equations. They are useful building blocks for
the study of the interaction of such waves in General Relativity.
A start in this direction has already been made with the description of
the head--on collision of the homogeneous plane waves of
section 2 in \cite{HW}.

The introduction of a cosmological constant is far more
complicated in the `plane wave' case than in the `spherical wave' case. We have
described this for the special case of adding a cosmological constant to high
frequency $pp$--waves. The introduction of a cosmological constant for the general
high frequency waves discussed in section 3 is a topic for further study.

\appendix
\section{Taking the limit $\lambda\rightarrow 0$} \setcounter{equation}{0}
The passage from eq.(\ref{1.3}) to eq.(\ref{1.7}) is brought about
quite simply as follows: for definiteness let us suppose that the
coordinate $u$ has a finite range $u_1\leq u\leq u_2$ during which
the radiation is detected. Let $u'$ be any value of $u$ within
this range and $\vartheta (u-u')$ be the Heaviside step function
which is equal to unity for $u-u'>0$ and which vanishes for
$u-u'<0$. This function acts as a convenient Greens' function for
the equation (\ref{1.3}). If we multiply (\ref{1.3}) by $\vartheta
(u-u')$ and integrate the equation with respect to $u$ in the
range $u_1\leq u\leq u_2$ then it is easy to see that the
resulting integral equation is
\begin{equation}\label{A1}
H_{\lambda}(u')=H_{\lambda}(u_2)+\int_{u_1}^{u_2}\left\{H_{\lambda}^2(u)+|W(u)|^2
\sin ^2\frac{u}{\lambda}\right\}\,\vartheta (u-u')\,du\
.\end{equation}We assume that all functions having a subscript
$\lambda$ have uniform $\lambda =0$ limits. If ${\cal
F}_{\lambda}(\zeta , \bar\zeta , u)$ is any such real--valued
function we assume that for small $\lambda
>0$ it has an expansion of the Isaacson \cite{I1} form:
\begin{equation}\label{A2}
{\cal F}_{\lambda}={\cal F}_0(\zeta , \bar\zeta ,
u)+\lambda\,f_1\left (\frac{u}{\lambda}\right )\,{\cal F}_1(\zeta
, \bar\zeta , u)+\lambda ^2\,f_2\left (\frac{u}{\lambda}\right
)\,{\cal F}_2(\zeta , \bar\zeta , u)+\dots\ ,\end{equation}where
$f_1, f_2$ etc. are real--valued functions. In order to take the
limit $\lambda\rightarrow 0$ of eq.(\ref{A1}) we make use of the
Riemann--Lebesgue theorem \cite{O} which states that if a
real--valued function $A(u)$ (it could depend on additional
variables such as $\zeta , \bar\zeta$, but its dependence on $u$
is the critical aspect for the theorem) is integrable (and
therefore could, for example, be a step function) on the interval
$u_1\leq u\leq u_2$ then
\begin{equation}\label{A3}
\lim_{\lambda\rightarrow
0}\int_{u_1}^{u_2}A(u)\,\sin\frac{u}{\lambda}\,du=0\
.\end{equation} It therefore follows that
\begin{equation}\label{A4}
\lim_{\lambda\rightarrow 0}\int_{u_1}^{u_2}A(u)\,\sin
^2\frac{u}{\lambda}\,du=\frac{1}{2}\,\int_{u_1}^{u_2}A(u)\,du\
.\end{equation}Thus taking the limit $\lambda\rightarrow 0$ of
(\ref{A1}) results in the integral equation
\begin{equation}\label{A5}
H_{0}(u')=H_{0}(u_2)+\int_{u_1}^{u_2}\left\{H_{0}^2(u)+\frac{1}{2}\,|W(u)|^2
\right\}\,\vartheta (u-u')\,du\ .\end{equation}Differentiating
this equation with respect to $u'$ (and using $d(\vartheta
(u-u'))/du'=\delta (u-u')$, the Dirac delta function) shows that
$H_0$ satisfies ({\ref{1.7})

The argument given here, when applied to the dependence of the
functions on the coordinate $u$, results in the passage from
(\ref{2.4}) to (\ref{2.6}). In addition writing (\ref{3.5}) in the
form
\begin{equation}\label{A6}
\frac{\partial}{\partial u}\left
(p_{\lambda}^{-3}m_{\lambda}\right
)-\frac{1}{4}\,p_{\lambda}^{-3}\Delta
_{\lambda}K_{\lambda}+p_{\lambda}\,|W|^2\sin
^2\frac{u}{\lambda}=0\ ,\end{equation} and applying the same
argument yields (\ref{3.7}).

\section{The Ricci Tensor Required for Section 4} \setcounter{equation}{0}
A convenient basis of 1--form fields for the space--time with line--element (\ref{4.1}) is
defined by:
\begin{eqnarray}\label{B1}
\vartheta ^1&=&P_{\lambda}^{-1}(d\zeta +\lambda\,P_{\lambda}^2\bar W\,\sin\frac{u}{\lambda}\,
d\bar\zeta )=-\vartheta _2\ ,\\
\vartheta ^2&=&\bar\vartheta ^1=-\vartheta _1\ ,\\
\vartheta ^3&=&B_{\lambda}\,du=\vartheta _4\ ,\\
\vartheta ^4&=&dr+\frac{1}{2}\,c_{\lambda}\,du=\vartheta _3\ .\end{eqnarray}
The components of the Ricci tensor calculated on the half--null tetrad given via these
1--forms are:
\begin{eqnarray}\label{B2}
R_{11}&=&\bar R_{22}=B_{\lambda}^{-1}\frac{\partial}{\partial\zeta}\left (P_{\lambda}^2
B_{\lambda}^{-1/2}\frac{\partial B_{\lambda}}{\partial\zeta}\right )+O(\lambda )\ ,\\
R_{12}&=&-\Delta _{\lambda}\log P_{\lambda}+\frac{1}{2}B_{\lambda}^{-1}\left (\Delta _{\lambda}
B_{\lambda}-B_{\lambda}^{-1}P_{\lambda}^2\left |\frac{\partial B_{\lambda}}{\partial\zeta}\right |^2
\right )+O(\lambda )\  ,\\
R_{13}&=&\bar R_{23}=-P_{\lambda}\,\frac{\partial}{\partial\zeta}\left (B_{\lambda}^{-1}
H_{\lambda}\right )++\frac{1}{2}B_{\lambda}^{-1}P_{\lambda}\,\frac{\partial ^2}{\partial u\partial\zeta}
(\log B_{\lambda})\nonumber\\
&&-\frac{1}{2}B_{\lambda}^{-1}P_{\lambda}\,\frac{\partial ^2c_{\lambda}}{\partial r\partial
\zeta}+O(\lambda )\ ,\\
R_{34}&=&-\frac{1}{2}B_{\lambda}^{-1}\Delta _{\lambda}B_{\lambda}-\frac{1}{2}B_{\lambda}^{-1}
\frac{\partial ^2c_{\lambda}}{\partial r^2}+O(\lambda )\ ,\\
R_{33}&=&-\frac{1}{2}B_{\lambda}^{-1}\Delta _{\lambda}c_{\lambda}-\frac{1}{2}P_{\lambda}^2
B_{\lambda}^{-2}\left\{\frac{\partial B_{\lambda}}{\partial\bar\zeta}\,\frac{\partial c_{\lambda}}{\partial\zeta}+
\frac{\partial B_{\lambda}}{\partial\zeta}\,\frac{\partial c_{\lambda}}{\partial\bar\zeta}\right\}\nonumber\\
&&-B_{\lambda}^{-2}H_{\lambda}\,\frac{\partial c_{\lambda}}{\partial r}-2\,B_{\lambda}^{-2}\dot H_{\lambda}
+2\,B_{\lambda}^{-2}H_{\lambda}^2\nonumber\\
&&+2\,B_{\lambda}^{-3}\dot B_{\lambda}\,H_{\lambda}+2\,B_{\lambda}^{-2}P_{\lambda}^4|W|^2\sin ^2\frac{u}{\lambda}
+O(\lambda )\ ,\\
R_{14}&=&\bar R_{24}=O(\lambda )\ ,\\
R_{44}&=&O(\lambda )\ .\end{eqnarray}Here as before
\begin{equation}\label{B3}
H_{\lambda}=P_{\lambda}^{-1}\dot P_{\lambda}\ ,\qquad \Delta _{\lambda}=2\,P_{\lambda}^2
\frac{\partial ^2}{\partial
\zeta\partial\bar\zeta}\ ,\end{equation}and the dot indicates partial differentiation with respect to $u$.

\section{The de Sitter Line--Element} \setcounter{equation}{0}
The form of the line--element of de Sitter space--time given in \cite{ORR} is of
paramount importance in the context of gravitational wave theory since it is based on a family
of null hyperplanes (generated by null geodesics with zero expansion and shear)
 in the space--time. Since it is well known (see for example \cite{S}) that the de Sitter
universe can be viewed as a quadric $V_4$ in 5--dimensional Minkowskian space--time $V_5$ we will
demonstrate how the Ozsv\'ath--Robinson--R\'ozga form of the de Sitter line--element
emerges from this point of view (this is discussed in a slightly different and separate form
in \cite{GDP}). The
line--element of $V_5$ is taken to be
\begin{equation}\label{C1}
-ds^2=\frac{3}{\Lambda}\,(dX^0)^2+(dX^1)^2+(dX^2)^2+(dX^3)^2-(dX^4)^2\
,\end{equation} and we assume that $\Lambda\neq 0$ ($\Lambda$ is the cosmological constant;
$V_4$ is a space--time of constant curvature $\Lambda /3$). The equation
of the quadric $V_4$ is given by
\begin{equation}\label{C2}
\frac{3}{\Lambda}\,(X^0)^2+(X^1)^2+(X^2)^2+(X^3)^2-(X^4)^2=\frac{3}{\Lambda}\
.\end{equation}The induced line--element on $V_4$ is got by
restricting (\ref{C1}) to $V_4$.

A useful shorthand for (\ref{C1}) and (\ref{C2}) is to write them
in an obvious vector notation \cite{S} as
\begin{equation}\label{C3}
-ds^2=d{\bf X}\cdot d{\bf X}\ ,\qquad {\bf X}\cdot {\bf
X}=\frac{3}{\Lambda}\ .\end{equation} Tran \cite{T} has shown
that the intersection of the quadric $V_4$ with the null
hyperplane passing through the origin of $V_5$,
\begin{equation}\label{C4}
{\bf b}\cdot {\bf X}=0\ ,\qquad {\bf b}\cdot {\bf b}=0\
,\end{equation}is a null hyperplane (whose geodesic generators have vanishing
expansion and shear) in $V_4$.

We can parametrise the points on $V_4\subset V_5$ using $(\zeta ,
\bar\zeta , r, u)$ by writing the position vector of a point on
$V_5$ in the form
\begin{equation}\label{C5}
{\bf X}={\bf Y}(\zeta , \bar\zeta , u)+p^{-1}q\,r\,{\bf a}(u)\
,\end{equation} with
\begin{eqnarray}\label{C6}
Y^0&=&p^{-1}\,\left (\frac{\Lambda}{6}\,\zeta\bar\zeta -1\right )\ ,\\
Y^1+iY^2&=&p^{-1}\sqrt{2}\,\zeta\ ,\\
Y^3&=&Y^4=p^{-1}\left\{\bar l(u)\,\zeta +l(u)\,\bar\zeta
+m(u)\,\left (1-\frac{\Lambda}{6}\,\zeta\bar\zeta\right )\right\}\
,\end{eqnarray}with $p=1+\frac{\Lambda}{6}\,\zeta\bar\zeta $,
\begin{eqnarray}\label{22}
a^0(u)&=&-\frac{\Lambda}{3}\,m(u)\ ,\\
a^1(u)+i\,a^2(u)&=&\sqrt{2}\,l(u)\ ,\\
a^3(u)-a^4(u)&=&-1\ ,\\
a^3(u)+a^4(u)&=&\frac{\Lambda}{3}\,m^2+2\,l\,\bar l\
,\end{eqnarray} and \begin{equation}\label{C7} q=\bar\beta
(u)\,\zeta +\beta (u)\,\bar\zeta +\alpha (u)\,\left
(1-\frac{\Lambda}{6}\,\zeta\bar\zeta\right )\ ,\end{equation} with
$\beta (u)=\dot l(u),\ \alpha (u)=\dot m(u)$. It follows from
these equations that
\begin{equation}\label{C8}
{\bf Y}\cdot {\bf Y}=\frac{3}{\Lambda}\ ,\qquad {\bf a}\cdot {\bf
Y}=0\ ,\qquad {\bf a}\cdot {\bf a}=0\ .\end{equation} Thus ${\bf
Y}$ is a point on $V_4$ corresponding to $r=0$ and, more
importantly, ${\bf X}$ in (\ref{C5}) is a point on $V_4$.
Substituting (\ref{C5}) into (\ref{C3}) gives the induced
line--element on de Sitter space due to its embedding in $V_5$.
This calculation is aided with the following scalar products:
\begin{equation}\label{C9}
{\bf a}\cdot \frac{\partial {\bf Y}}{\partial u}=-p^{-1}q\ ,\qquad
\frac{\partial {\bf Y}}{\partial u}\cdot\frac{\partial {\bf
Y}}{\partial u}=0\ ,\qquad \dot {\bf a}\cdot\frac{\partial {\bf
Y}}{\partial u}=0\ .\end{equation}We also have $\dot {\bf a}\cdot
\dot {\bf a}=\frac{\Lambda}{3}\,\alpha ^2+2\,\beta\bar\beta$ and
the further scalar products:
\begin{equation}\label{C10}
\frac{\partial {\bf Y}}{\partial\zeta}\cdot\frac{\partial {\bf
Y}}{\partial\bar\zeta}=p^{-2}\ ,\ \frac{\partial {\bf
Y}}{\partial\zeta}\cdot\frac{\partial {\bf
Y}}{\partial\zeta}=0=\frac{\partial {\bf
Y}}{\partial\bar\zeta}\cdot\frac{\partial {\bf
Y}}{\partial\bar\zeta}\ ,\ \dot {\bf a}\cdot\frac{\partial {\bf
Y}}{\partial\zeta}=\frac{\partial}{\partial\zeta}(p^{-1}q)\
.\end{equation}This calculation results in the line--element
\begin{equation}\label{C11}
ds^2=-2\,p^{-2}d\zeta\,d\bar\zeta
+2\,p^{-2}q^2du\,dr+p^{-2}q^2(2\,q^{-1}\dot q\,r-\kappa (u)\,r^2)\,du^2\ ,\end{equation}
with $\kappa (u)=2\,\beta\,\bar\beta +\Lambda\,\alpha ^2/3$,
which is the line--element of de Sitter space in the interesting form given by Ozsv\'ath, Robinson and
R\'ozga \cite{ORR}. Using the construction of de Sitter space as the quadric
(\ref{C2}), Tran \cite{T} has provided beautiful proofs that properties of the functions
$\alpha (u) , \beta (u)$ and $\bar\beta (u)$, along with the sign of the cosmological
constant $\Lambda$, determine whether or not the null hypersurfaces $u={\rm constant}$
intersect [Tran's results \cite{T} are summarized as follows:
if $\Lambda >0$ then $\kappa >0$ implies intersections while
if $\Lambda <0$ then (a) $\kappa <0$ implies intersections, (b) $\kappa =0$
implies intersections except if ${\rm Im}\beta =0$ or ${\rm Re}\beta =0$
or ${\rm Re}\beta =C\,{\rm Im}\beta$ for some constant $C$ and (c) $\kappa >0$ implies
intersections].

\end{document}